# Characterization and collection of information from heterogeneous multimedia sources with users' parameters for decision support

Robert Charles
Laboratoire Lorrain de recherche en Informatiques et ses Applications,
LORIA Campus scientifique,  BP 239 54506, Vandoeuvre les Nancy, France
charles.robert@robert-scientist.com

**ABSTRACT**
No single information source can be good enough to satisfy the divergent and dynamic needs of users all the time. Integrating information from divergent sources can be a solution to deficiencies in information content. We present how Information from multimedia document can be collected based on associating a generic database to a federated database. Information collected in this way is brought into relevance by integrating the parameters of usage and user's parameter for decision making. We identified seven different classifications of multimedia document.

## 1. INTRODUCTION

Multimedia is defined as "*the combination of various presentation media such as text, sound, graphics, animation, and video*"[1]. Multimedia information spans across cultural boundaries, domain of activities and formats of presentation. In as well as we are not interested in examining forms of multimedia information, we are content to say that, they can be very interesting as source of information. We identify multimedia documents as document containing more than one type of support for document. In attempting to evaluate multimedia document, it was noted that a mirage of grouping and classifications exist based on content or on the structure of the document. It is not our intention to approve or disapprove these classifications and assertions.

## 2. CONCEPTIONS IN INFORMATION AGGREGATION

We believe that a document is a container of information. A document is expected to transmit information. This means that, a multimedia information management system is the management of content of multimedia documents. We assume that most objects that can be referred to as document has one or more characteristics of a multimedia document. A multimedia document can be defined as any document having one or more of the following: Text, sound and image. In the table 1, we give example

Several approaches have been attempted in information extraction from multimedia document. Some of these approaches are based on imposing additional information layer based on human or machine translation of multimedia document. For example, texts information can be derived from video streams or images [3]. It must be emphasized that these approaches are not necessarily information extraction but information creation that are subjective to human or machine understanding of events. Sometime these interpretations may be divergent view of what the author of the document intended.

---
[1] http://www.scala.com/definition/multimedia.html

**Table 1**: Possible forms of multimedia documents

|   | Text | Sound | Image | Example |
|---|------|-------|-------|---------|
| 1 |      |       | X     | Paints  |
| 2 |      | X     |       | Music   |
| 3 |      | X     | X     | Video   |
| 4 | X    |       |       | Book    |
| 5 | X    |       | X     | Commented image |
| 6 | X    | X     |       | Advertisement |
| 7 | X    | X     | X     | Commented Video |

In a related work [8], information extraction was approached by identifying three separate area poles. These poles were reported as (a) document Information: global information about each document including meta-information and raw-data information. (b) document structure: the temporal decomposition of video documents that comes from the temporal segments covered by the description data and (c) document description: the set of description data that is either automatically extracted (feature-based) or entered manually by human operators. This work can be viewed from viewpoint of information creation and not information extraction or information use. It may be assumed that information extraction from multimedia document integrates creating a layer of additional information for its classification and usage. It is not to the knowledge of the author if there are systems that can "interpret" multimedia documents and generate meaningful information embedded there in.

Existing work in federated database management proposed two models of information collection in a federated system presented. These are schema integration and schema coordination approach to cooperative query processing [11]. This approach is well suited to databases or meta-databases that are homogeneous in their format. In the case of heterogeneous multimedia document, a different approach is expected particularly when interest is for decision making.

The conception in this work is that all multimedia documents can be classified in one of the category in Table 1. The word multimedia means "more than one", but in this case, it was believed that multimedia information source can include apparently one of the three characteristics. This is because, the terms "text", "image" and "sound" needed a further clarification that is beyond the objective of this work. How image, text or sound be defined? What are the criteria for defining these terms? Definition of image for instance is subjective. Looking at some artistic writings, can they be called text or text and image? For this purpose, the frontiers of separation between the three isolated states of these elements remain unclear.

We represented multimedia documents in a way that is not exclusive but a guide. Some additional information may be included when necessary. Each of the class of the identified multimedia document is described and qualified in the following section. The starting point is the Text-Image-Sound because presumably, it may contain all the possible characteristics of a multimedia document.

**Text-Image-Sound:** We are paying attention to representing this class of multimedia document as much as possible since it comprise all possible form of multimedia document. Our representation here is in three groups of text, image and sound.

In order to reduce the complexity of documents and make them easier to handle, attempts have been made by transforming them from full text version to a document vector thereby describing the contents of the document [7] or by approach of vector space representation [10]. In most cases, representation of textual document is directly associated to the use of such representation.

Borrowing from information science, computer science and Dublin Core initiatives, this work identified text multimedia document as a sub-set of text-image-sound was identified with the following characteristics the study:
- author
- title
- summary
- reference date: the reference date may refer to the date in which the entire document associated to the text was published
- descriptors
- key words
- related document(s)

In order to understand images, a work [5] emphasised the dependency of the external information in an image on its structural composition. It emphasised the perspective of high-level and low-level compositions. In [4], it was proved that object recognition is dependent on the understanding of the concepts of low-level and high-level composition of images. The importance of images in knowledge representation was highlighted by [1]. In a clear term, it presented a semantic representation of images applied to knowledge and reasoning. It is possible to represent images from different perspectives like low-level, high-level and structural level. It was assumed that a low-level representation of images based on the pixel value of segments is not enough in a decision support environment. For example how can the RGB or CMYK colour scheme values of pixels be reconciled with decisional problems? Despite the tremendous importance that this kind of representation may have particularly in the area of digital image processing, the interest is in the specificity of images shown in their physical properties (visible properties) such as:
- dominance colour (red, orange, yellow, green, blue, indigo, violet, grey, black, white, etc)
- secondary colour (red, orange, yellow, green, blue, indigo, violet, grey, black, white, etc)
- dominant shape (oval, circle, square, rectangle, triangle, cylindrical, rhombus, irregular, line)
- secondary shape (same as in dominant shape)
- specificity of shape (repeated, perfect shape, deformed, interposed,….)
- dominant object (equipment, tool, )
- specificity of object (deformed, at foreground, at background, …)
- secondary object,

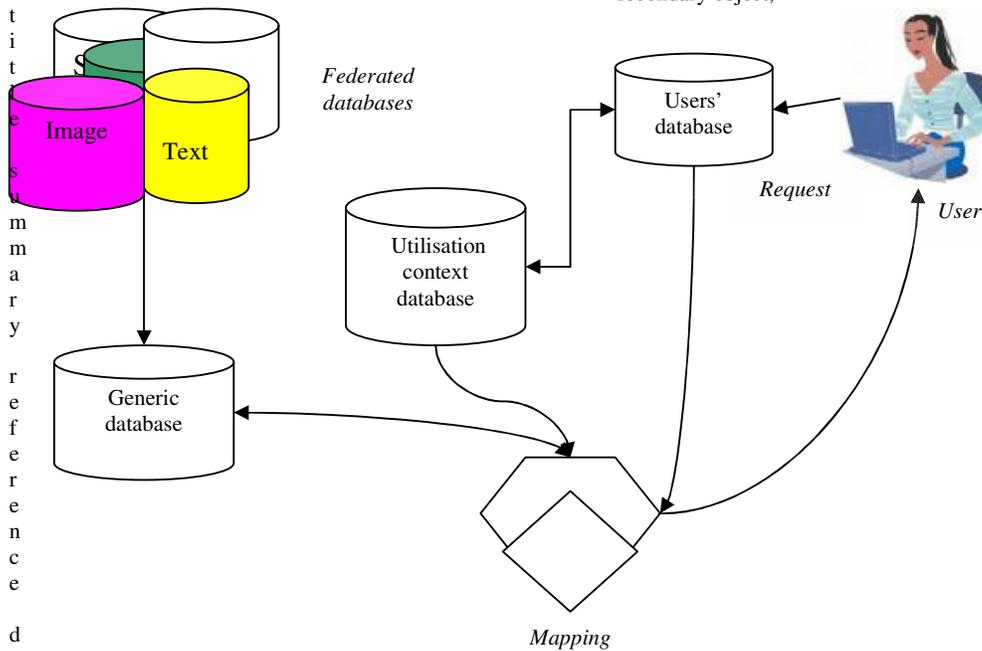

**Figure. 1:** Aggregating information with user's consideration

- dominant feature (nature, water body, sporting, animal, human being, activity)
- secondary feature,
- sub-class of dominant feature (animal-mammal, animal-wild, animal-domestic, water-ocean, activity-war, activity-manufacturing…..)
- sub-class of secondary feature
- image format
- image medium (wood, electronic, paper, glass, stone, plastic, composite, etc.)
- image type (water colour, digital image, oil colour, sketch, humour, cartoon, etc).

One method used in representing sound was based on the idea of "spectro-temporal" fields response and the use of convolution to measure the degree of similarity between the feature detectors and the stimulus [2].

We consider sound with the following identities:
- originator,
- target (public, private, not specified)
- descriptors,
- date of publication,
- sound type (noise, music or voice),
- sound class (debate, dialogue, music, publicity, …)
- sound sub-class (country music, blast noise, industrial noise, warning sound, disorder…)

**Text:** This is the commonest type of edited information. Written textual information take different forms of supports such as news paper, book, electronic information (web site). Sometime textual information may be found on supports such as stones, building or even on animal or plants.

**Image:** The term "image" takes its origin from Latin "imago" which implies "designed for visual representation". They are multimedia document presented using different support such as painting, design, photographs, digital image, etc. Different categories of images in a document can be cited. For example design, maps, musical notes. For this study, it was not directed to structural composition of an image but their physical properties.

**Sound:** Sound was considered as a type of multimedia document that appeal to audio sensation of man. Three types of sound were identified as noise, music and voice. There was no attempt made to differentiate this subclass because the classification is subjective. What was considered as noise under a certain circumstance can be seen as music in another circumstance. The concern was not with specific technical characteristics of these qualifications.

**Text - Image:** We define multimedia document of type "text-image" as multimedia document that has the characteristics of text and the same time characteristics of an image.

**Text-Sound:** A representation comprising all the characteristics of text and that of sound.

**Image-Sound –:** We classify assume image-sound as a multimedia document combining properties of document containing only sound and document containing only image. They are multimedia document that can not be described with only sound or only image but the two.

## 3. OUR PROPOSAL

Though we have given a generalized but specific representation of multimedia documents, it was thought that, it is near impossible to represent all the possible format of multimedia information sources. The description of the attempt and proposition for collection of information coming from heterogeneous multimedia sources can be explained with the Figure 1. It consists of two databases: generic (derived database) and federated database. The federated databases consist of all imaginable databases (data bank) of all formats. Each of the databases has its own independent specificity. They are mutually independent. We may have database that consist of text whereas some may contain a mixture of text and sound. It is difficult for us to imagine the processes that may be involved in "synchronizing" specific database to a generalized database as we proposed. It is quite difficult to imagine the variety in formatting, medium, and content. This calls for a profound consideration that may permit the assessment of "interface" between federated databases and derived database. In short, detail information in federated database is not of great importance to us because of its complexity.

In the case of derived database, we have a general representation of every possible form of multimedia document impliedly represented as Text-Image-Sound in the previous section.

The animation property of a multimedia document was not particularly of interest because animation property will involve much more complex characterization that is beyond the scope of this work. Questions that are to be considered in such cases include: How can the animation in a multimedia document be represented? It was assumed that animation of a multimedia document is nothing but changes in the location of a document with time. For example, a document or part of a document can be said to be found at X,Y,Z at time $T_1$ while at time $T_2$, the location may be said to be found at X+X$dx$, X+Y$dy$, Z+Z$dz$. In fact, the animation characteristic of a document requires other interpretation and processing for additional information in the document.

From table 1, it was assumed there can be seven types of multimedia information sources. It was presupposed that these seven types are implied in federated multimedia sources. Federated because, a satellite database (derived database) that is directly linked to the federated sources can be created. Adequate information was stored and retrieved from derived database. When necessary, fuller information can be requested from host sources in the federated database.

**Generic database:** The objective of creating a generic database is not to provide comprehensive information for use in decision making but as an extract from federated database and as a link to information sources. Generic database was a reflection of document source that can include any of the seven classes of multimedia information type presented in *table 1*. This database should contain enough information as to facilitate its independent usage. In this study, we are not particularly interested in the complexity of database design to avoid redundancy or data integrity. The objective was to have as much information as possible in the generic database to allow for its independent use without necessarily referring to the complete document in the federated database. Among other parameters, it was contemplated that a generic database (derived database) should contain the following:

**Document code:** The document code is a reference relating the generic database to the federated database where complete information in respect to specific item can be obtained. Document code may as well be an internet link.

**Text-Image-Sound:** All the characteristics of text, image and sound in the federated database.

It was believed that a generic database can not be completely useful without integrating usage parameter into the source database. The usage parameter other wise called context of usage was seen from two perspectives. There are dynamic part and the static contexts of usage. The static context of usage of multimedia database was defined from historical point of view. In other words, the concern was from the view of what has been done (repeatedly) with multimedia database. The work of [6] specified four usage contexts as regards multimedia databases. The usage contexts are (a) Teaching (b) Learning (c) documentation and (d) entertainment. There was no attempt to explain in detail what these usages are other than taking the meaning from their facial level. The representation of context of usage was with following parameters:

- Date of usage
- Context of usage (existing usage, new usage)
- User's identity
- Generic document identity
- Type of use (repetitive use, occasional)

If multimedia database have been used in these four contexts, it implied that men have used multimedia database in four major ways. It was noted that as social being, human need is not static therefore context of use of multimedia database may not necessarily be static. In a dynamic context of use of multimedia database, we are interested in enriching our context of usage database with new usages. This calls for the mapping of human factor (or user's profile in users' database) into the context of use. It is therefore obvious that the class of usage in context of usage may not be the same all the time.

Table 2: Cross analysis of information use based on parameters of use

|   |   | Parameters of usage | | | | Representation |
|---|---|---|---|---|---|---|
|   |   | Doc | Context | User | Time |   |
| 1 | The complete context of usage of all document by all users all time |   |   |   |   | $\iiiint dDdCdUdT$ |
| 2 | The users of all document in all context at a time |   |   |   | X | $T\iiint dDdCdU$ |
| 3 | Usage habit of a user |   |   | X |   | $U\iiint dDdCdT$ |
| 4 | Usage habit of a user at a specific time |   |   | X | X | $UT\iint dDdC$ |
| 5 | Documents used in a specific context and their users |   | X |   |   | $C\iiint dDdUdT$ |
| 6 | Documents used in a specific context and their users at a specific time |   | X |   | X | $CT\iint dDdU$ |
| 7 | Which documents were used by a user a specific context |   | X | X |   | $CU\iint dDdT$ |
| 8 | Which documents were used by a user a specific context at a specific time |   | X | X | X | $CUT\int dD$ |
| 9 | How a document was used | X |   |   |   | $D\iiint dCdUdT$ |
| 10 | How a document was used at a specific time and the users | X |   |   | X | $DT\iint dCdU$ |
| 11 | How a document was used by a user | X |   | X |   | $DU\iint dCdT$ |
| 12 | How a document was used by a user at a specific time | X |   | X | X | $DUT\int dC$ |
| 13 | Who are the users that used a document | X | X |   |   | $DC\iint dUdT$ |
| 14 | Who are the users that used a document at a specific time | X | X |   | X | $DCT\int dU$ |
| 15 | How many time was a document used in a specific context by a user | X | X | X |   | $DCU\int dT$ |
| 16 | Why was a document used in a specific context by a user at a time | X | X | X | X | $DCUT$ |

The user's database may contain the following elements
- User's identity
- User's name
- User's address
- Social class

In the case of the federated database, we present some of the characteristics that will be of interest to us in table 2:

## 4. ANALYSIS

We used four parameters distinctive from the works in competitive intelligence and decision sciences to characterize decision making processes. It was believed that the four parameters used in representation of a multimedia usage can be used in the context of decision making. Some of the information that can be derived from these four parameters is represented in table 1.

It may be interesting to know all the possible ways a particular multimedia document has been used ($D\iiint dCdUdT$) to enable the reclassification of a document or to be able to know the importance of a document. It may be of interest to know which documents were used by a particular users ($CU\iint dDdT$) to enable us determine his research interest for example. The various ways multimedia documents are being used ($\iiiint dDdCdUdT$) may be of interest to enable the knowledge of the evolution of document usage. We earlier pointed out the fact that context of usage of document is dynamic.

Presentation of detailed possible analysis may not be necessary since it is possible to identify complete intra-parameter or sub-parameters evaluation. For example, the interest of knowing how many users use documents occasionally as compared to repetitive use of documents may be at stake. Comparison of frequency of context usages against user's social class is possible.

## 5. CONCLUSION AND PERSPECTIVE

Demonstration of the fact that every multimedia document can effectively be represented with some or all the characteristics of a document of type text-image-sound type was presented. The proposal was of two levels of databases (federated databases and generic database). The attempt was to clarify that the context of use of document was of two types which qualified it for decision making. Examples of information that can be derived from such approach were given.

Information system based this proposal is being implemented on internet by integrating information from all kinds of information bases borrowing from conceptions of "meta-search engine" systems such as metacrawler.com. A generic database is being put in place on a local server.

## 6. REFERENCES

[1.] Bloehdorn S., Petridis K., Simou N., Tzouvaras V., Avrithis Y., Handschuh S., Kompatsiaris Y., Staab S. and Strintzis M. G., Knowledge Representation for Semantic Multimedia Content Analysis and Reasoning, Proceedings of the European Workshop on the Integration of Knowledge, Semantics and Digital Media Technology (EWIMT). November 2004, (2004).

[2.] Coath M. and Denham S. L., Robust sound classification through the representation of similarity using response fields derived from stimuli during early experience, Biological cybernetics, (2005) Jul;93(1):22-30. Epub 2005 Jun 8

[3.] Duygulu, P. and Wactlar, H., Associating Video Frames with Text, 26th Annual International ACM SIGIR Conference, July 28-August 1, (2003),Toronto, Canada.

[4.] Lindgren J.T. and Hyvärinen A., Learning high-level independent components of images though a spectral representation, Proceedings of the 17th International Conference on Publication Date: 23-26 Aug. (2004), Volume: 2, pages 72- 75 Vol.2

[5.] Macieszczak M. and Ahmad O., On the polygonal decomposition in low-level processing for image understanding, Canadian Conference on Electrical and Computer Engineering, (1994). Conference Proceedings. Page(s): 457-460 vol.2

[6.] Maghrebi Hanene and David Amos, Toward a model for the representation of multimedia information based on users' needs: economic intelligence approach, IV International conference on Multimedia and Information and Communication Technologies in Education, m-ICTE2006. 22-25 November, (2006) Seville, Spain.

[7.] Meadow, C. T. (1992), Text Information Retrieval Systems. Academic Press,.

[8.] Moënne-Loccoz N., Bruno J., Marchand-Maillet S., and Bruno E., An Integrated Framework for the Management of Video Collection, in MLMI 2004, LNCS 3361, pp. 101–110, 2005, Springer-Verlag Berlin Heidelberg, (2005),

[9.] Robert C. and David A., Annotation and its application to information research in economic intelligence, Advances in Knowledge Organization, Vol 10, pages 35-39, (2006), Ergon Verlag, Wuzburg, Germany

[10.] Xiaofei H., Deng C., Haifeng L. and Wei-Ying M., Locality Preserving Indexing for Document Representation, SIGIR'04, July 25–29, (2004), Sheffield, South Yorkshire, UK.

[11.] Zhao J. Leon, Schema coordination in federated database management: A comparison with Schema integration, Workshop on Information Technology and Systems, December 17-18, (1994), Vancouver, Canada, URL: http://citeseer.ist.psu.edu/204504.html (28/11/2006)